\documentclass[aps,prl,twocolumn,showpacs,preprintnumbers,amsmath,amssymb,
epsf,superscriptaddress,groupedaddress, tightenlines]{revtex4}

\usepackage{graphicx}
\usepackage{amsmath}
\usepackage{bm}
\newcommand{\nn}{\nonumber}
\newcommand{\vslash}{v\hspace*{-5.5pt}\slash}

\newcommand{\Bslash}{{\cal B}\hspace*{-5.5pt}\slash}
\newcommand{\Dslash}{D\hspace*{-5.5pt}\slash}
\newcommand{\varepsslash}{\varepsilon\hspace*{-5.5pt}\slash}
\newcommand{\lqcd}{\Lambda_{QCD}}
\newcommand{\Dleft}{\overleftarrow D}
\def\nslash{n\!\!\!\slash}
\def\bnslash{\bar n\!\!\!\slash}
\def\bn{\bar n}

\begin{document}
\preprint{\vbox{\hbox{UCSD/PTH 05--1}\hbox{MIT-CTP 3592}}}

\title{
Chiral symmetry and exclusive  B decays in the SCET}



\author{Benjam\'\i{}n Grinstein}
\affiliation{Department of Physics, University of California at San Diego,
  La Jolla, CA 92093}

\author{Dan Pirjol}
\affiliation{Center for Theoretical Physics, MIT, Cambridge, MA 02139}


\date{\today}

\begin{abstract}
We construct a chiral formalism for processes involving
both energetic hadrons and soft Goldstone bosons, which extends the
application of soft-collinear effective theory to multibody B decays.
The nonfactorizable helicity amplitudes for heavy meson decays into multibody
final states satisfy symmetry relations analogous to the large energy
form factor relations, which are broken at leading
order in $\Lambda/m_b$ by calculable factorizable terms. 
We use the chiral effective theory to compute the leading corrections
to these symmetry relations in $B\to M_n \pi \ell\bar \nu$ and
$B\to M_n \pi \ell^+\ell^-$ decays.
\end{abstract}



\maketitle


1. \textit{Introduction.} 
The study of processes involving energetic quarks and gluons
is simplified greatly by going over to an effective theory
which separates the relevant energy scales. The soft-collinear
effective theory (SCET) \cite{Bauer:2000ew} simplifies the proof of
factorization theorems and allows a systematic treatment of
power corrections. SCET has been applied to both inclusive and
exclusive hard processes with energetic final state particles.

In this paper we present a combined application of the SCET with chiral 
perturbation theory which can be used to study exclusive processes 
involving both energetic light hadrons and soft pseudo Goldstone bosons and
photons. The main observation is that once the dynamics of the 
collinear degrees of freedom has been factorized from that of the soft
modes, usual chiral perturbation theory methods can be applied to the latter,
unhampered by the presence of the energetic collinear particles which
might have upset the momentum power counting in $p/\Lambda_\chi$.
The chiral formalism has been applied previously to compute matrix elements
of operators appearing in hard scattering processes, such as DIS and
DVCS \cite{parton,parton1,ChSa}. Our paper extends these results to 
processes with both soft and collinear hadrons.

We focus here on exclusive B decays, which are described by 
three well-separated scales: hard $Q\sim m_b$, hard-collinear
$\sqrt{\Lambda Q}$ and the QCD scale $\Lambda \sim 500$ MeV. 
This requires the introduction of a sequence of
effective theories QCD $\to$ SCET$_{\rm I} \to$ SCET$_{\rm II}$, containing
degrees of freedom of successively lower virtuality \cite{bpsff}.
The intermediate theory ${\rm SCET}_{\rm I}$ contains hard-collinear quarks
$\xi_n$ and gluons $A_n^\mu$ with virtuality $p_{\rm hc}^2 \sim \Lambda Q$ 
and ultrasoft quarks and gluons $q, A_\mu$ with virtuality $\Lambda^2$.
Finally, one matches onto ${\rm SCET}_{\rm II}$ which includes only
soft $q, A_\mu$ and collinear $\xi_n, A_n^\mu$ modes with virtuality 
$p^2 \sim \Lambda^2$.
The expansion parameter in both effective theories can be chosen
as $\lambda^2 \sim \Lambda/m_b$.

In the low energy theory SCET$_{II}$ the soft and collinear modes
decouple at leading order and the effective Lagrangian is simply a
sum of the kinetic terms for each mode
\begin{eqnarray}
{\cal L}^{(0)} = {\cal L}_\xi^{(0)} + \sum_q \bar q (i\Dslash-m_q) q
+ {\cal L}_{A_n}^{(0)} \,.
\end{eqnarray}
The matching of an arbitrary operator $O$
onto SCET$_{II}$ can be written symbolically as \cite{bpsff}
\begin{eqnarray}\label{fact}
O \to T \otimes O_S \otimes O_C + O_{\rm nf} + \cdots
\end{eqnarray}
where the ellipses denote power suppressed contributions. 
The first term is a `factorizable' contribution, with $O_S, O_C$
soft and collinear operators convolved with a Wilson coefficient
$T$ depending on the arguments of $O_S, O_C$.
$O_{\rm nf}$ denote `nonfactorizable' operators. Their
precise form depends on the IR regulator adopted for SCET$_{\rm II}$;
for example, in dimensional regularization they might take the form
of $T$ products of operators involving messenger modes \cite{mess}. 

This formalism has been used to study exclusive B decays into
energetic light hadrons (e.g. $B\to \pi\ell\nu$ and $B\to K^*\gamma$)
\cite{bpsff,Beneke:2003pa,CK,pol},
and nonleptonic decays into 2 energetic light hadrons such as
$B\to \pi\pi$ \cite{CKpipi,bprs}. This paper presents an extension
of this formalism to describe multi-body B decays to one energetic
hadron plus multiple soft pions and photons. Such decays received increased 
attention recently \cite{multi,exp} due to their ability to extend the
reach of existing methods for determining weak parameters.

In Sec.~2 we introduce the SCET formalism and review the derivation of the
large energy symmetry 
relations for the $B\to M$ form factors \cite{ffrel,bpsff,ps1}. We show that similar
relations exist for B decays into multibody final states containing one
collinear hadron $M_n$ plus soft hadrons $X_S$, $B\to M_n X_S$.
Sec.~3 develops a chiral formalism for computing the matrix elements of the
soft operators in (\ref{fact}) $\langle X_S|O_S|B\rangle$
with $X_S$ containing only soft Goldstone bosons. 
As an application we discuss in Sec.~4 the semileptonic and rare radiative 
decays  $B\to M_n \pi_S \ell\bar\nu$ and $B\to M_n \pi_S \ell^+\ell^-$.
\vspace{0.5cm}

2. \textit{Symmetry relations.} 
The most general SCET$_{I}$ operator appearing in the matching of
SM currents $\bar q\Gamma b$ for $b\to u\ell\bar \nu$ or $b\to s\gamma$ decays has
the form (we neglect here light quark masses, which can be included as in
\cite{masses})
\begin{eqnarray}\nn
 J^{\rm eff}_\mu &=& 
  c_1(\omega)\, \bar q_{n,\omega} \gamma^\perp_\mu P_L\, b_v\\
&& +\,
[c_2(\omega) v_\mu +  c_3(\omega) n_\mu ]\,\, \bar q_{n,\omega} P_R\, b_v \\
\label{JeffSCET}
&+&  b_{1L}(\omega_i)\, J^{(1L)}_{\mu}(\omega_i) 
  + b_{1R}(\omega_i)\, J^{(1R)}_\mu(\omega_i) \nn \\
&& + \, [b_{1v}(\omega_i) v_\mu + b_{1n}(\omega_i) n_\mu ]\,\, J^{(10)}(\omega_i)  \nn
\end{eqnarray}
These are the most general operators allowed by power counting and which 
contain a left-handed collinear quark. We neglect $O(\lambda)$ operators of
the form $\bar q_n {\cal P}_\perp^\dagger \Gamma b_v$ which do not contribute
below. The relevant modes
are soft quarks and gluons with momenta $k_s \sim \Lambda$ and 
collinear quarks and gluons moving along $n$.
$n_\mu,\bar n_\mu$ are unit light-cone vectors satisfying
$n^2 = \bar n^2 = 0, n\cdot \bar n = 2$.

The $O(\lambda)$ operators are defined as
\begin{eqnarray}
J^{(1L,1R)}_\mu(\omega_1,\omega_2) &=& \bar q_{n,\omega_1}\, 
\Gamma^{(1L,1R)}_{\mu\alpha}
  \Big[\frac1{\bar n\cdot {\cal P}} ig {\cal B}^\alpha_{\perp n}\Big]_{\omega_2}
  b_v , \nn\\
J^{(10)}(\omega_1,\omega_2) &=& \bar q_{n,\omega_1}\, 
  \Big[\frac1{\bar n\cdot {\cal P}} ig \Bslash^\perp_n\Big]_{\omega_2}
  P_L\, b_v , 
\end{eqnarray}
with $\{\Gamma^{(1L)}_{\mu\alpha}\,, \Gamma^{(1R)}_{\mu\alpha}\} =
\{\gamma^\perp_\mu \gamma^\perp_\alpha P_R\,, \gamma^\perp_\alpha \gamma^\perp_\mu P_R\}$.
The action of the collinear derivative $i\partial_\mu$ on collinear
fields is given by the momentum label operator 
${\cal P}_\mu = \frac12 n_\mu \bar n\cdot {\cal P} + {\cal P}^\perp_\mu$.
The collinear gluon field tensor is $ig {\cal B}_\mu = W^\dagger
[\bar n\cdot iD_c, iD_{c\mu}^\perp] W$.
The Wilson coefficients $c_i,b_i$ depend on the Dirac structure of the QCD current
$\Gamma$ and are presently known to next-to-leading order in matching
\cite{Bauer:2000ew,oneloop1,oneloop2}.

After matching onto SCET$_{\rm II}$, the effective current
(\ref{JeffSCET}) contains the factorizable operators
\begin{eqnarray}\label{fact2}
&& J_\mu^{\rm fact} =  -\frac{1}{2\omega} 
\int dx dz dk_+ b_{1L}(x,z) J_\perp(x,z,k_+) \\
&&\qquad\qquad \times 
((\bar qY)_{k_+} \nslash \gamma^\perp_\mu \gamma_\perp^\lambda P_R (Y^\dagger b_v))
(\bar q_{n,\omega_1} 
\frac{\bnslash}{2} \gamma_\perp^\lambda q_{n,\omega_2}) \nn\\
&&
- \frac{1}{2\omega}
\int dx dz dk_+ b_{1R}(x,z) J_\parallel(x,z,k_+) \nn \\
&&\qquad\qquad \times 
((\bar qY)_{k_+} \nslash \gamma^\perp_\mu P_R (Y^\dagger b_v))(\bar q_{n,\omega_1} 
\frac{\bnslash}{2} P_L q_{n,\omega_2}) \nn \\
&&
- \frac{1}{\omega}
\int dx dz dk_+ [b_{1v}(x,z) v_\mu + b_{1n}(x,z) n_\mu)]\nn\\
&& \qquad \times J_\parallel(x,z,k_+)
 ((\bar qY)_{k_+} \nslash  P_L (Y^\dagger b_v))(\bar s_{n,\omega_1} 
\frac{\bnslash}{2} P_L q_{n,\omega_2}) \nn 
\end{eqnarray}
where $J_{\perp,\parallel}$ are jet functions defined as in \cite{bprs}.
We denoted here $\omega_1 = x\omega, \omega_2 = -\omega (1-x), 
\omega = \omega_1 - \omega_2$.
This has the factorized form of Eq.~(\ref{fact}), with the Wilson 
coefficient $T$ given by  $b_i \otimes J_{\parallel,\perp}$.

\begin{table}
\caption{\label{table} Counting the independent hadronic
parameters required for a general
$B\to M_n X_S$ decays in QCD, SCET and for a 1-body hadronic state $X_S=0$.}
\begin{ruledtabular}
\begin{tabular}{c|c|c|c}
 & constraints & parameters & \# of indep. \\
 &             &            & parameters \\
\hline
QCD & -- & $H^{V-A}_{\pm,L,0}\,,H^{T}_{\pm,L}$ & 7 \\
SCET & $H^{V-A}_\lambda \propto H^{T}_\lambda\,, H_0\propto H_t$ & 
$\zeta_{\perp,0}\,, S^{(L,R)}$ & 2 + 2 \\
1-body & $H_+^{V-A} \,, H_+^T \sim O(\frac{\Lambda}{m_b})\,, $ & 
$\zeta_{\perp,0}\,, S^{(0)}$ & 2 + 1 \\
\end{tabular}
\end{ruledtabular}
\end{table}

The nonfactorizable operator $O_{\rm nf}$ in Eq.~(\ref{fact}) arises from 
matching the LO SCET$_{I}$ operators onto SCET$_{II}$ \cite{bpsff}. 
The precise form of the latter operators is not essential for our argument,
which depends only on the Dirac structure of the SCET$_{I}$ operators.
Before proceeding to write down the SCET predictions for these matrix elements,
we define more precisely the kinematics of the process.

The transition $B\to M_n X_S$ induced by the current $J_\mu = \bar q\Gamma_\mu b$ can
be parameterized in QCD in terms of 4 helicity amplitudes defined as
\begin{eqnarray}
H_\lambda^{(\Gamma)}(M_n,X_S) = \langle M_n X_S|\bar q \Gamma_\mu \varepsilon^{*\mu}_\lambda 
b| B\rangle
\end{eqnarray}
with $\varepsilon_{\pm, 0, t}^\mu$ a set of four orthogonal unit vectors 
defined in the rest frame of $v$ as
$\varepsilon_\pm^\mu = \frac{1}{\sqrt2}(0,1,\mp i,0)\,, \varepsilon_0^\mu = 
\frac{1}{\sqrt{q^2}}
(|\vec q|,0,0,q_0)\,, \varepsilon_t^\mu = \frac{1}{\sqrt{q^2}} (q_0, 0, 0, |\vec q|)$.
These definitions correspond to the choice $n=(1,0,0,1)$, $\bar n = (1,0,0,-1)$.

In the language of helicity amplitudes, 
the most general matrix elements of the nonfactorizable operators
are given in terms of the 2 parameters
\begin{eqnarray}\label{zeta}
&&
\langle M_n X_S|\bar q_{n,\omega} \varepsslash_-^* P_L b_v|\bar B\rangle = 
2E_M \zeta_\perp(E_M,X_S)\\
&& \langle M_n X_S|\bar q_{n,\omega} P_R b_v|\bar B\rangle = 2E_M \zeta_0(E_M,X_S)
\nn
\end{eqnarray}
$\zeta_{\perp,0}(E_M,X_S)$ are complex quantities depending on the momenta, spins 
and flavor of the particles in the final state.

The relations Eq.~(\ref{zeta}) imply several types of SCET predictions for 
the nonfactorizable
contributions to the helicity amplitudes.
The most important one is the vanishing of the right-handed (nonfactorizable)
helicity amplitudes at leading order in $1/m_b$, for any current $\Gamma$ 
coupling only to left chiral collinear quarks
\begin{eqnarray}\label{hel0}
H_+^{\rm nf}(\bar B \to M_n X_S) = 0\,.
\end{eqnarray}
For decays to one-body states, this constraint leads to the well-known 
large energy form factor relations $m_B/(m_B+m_V) V(E)= (m_B+m_V)/(2E) A_1(E)$ 
(for $\Gamma_{V-A} = \gamma_\mu P_L$) and 
$T_1(E) = m_B/(2E) T_2(E)$ (for $\Gamma_T = i\sigma_{\mu\nu} q^\nu P_R$)
\cite{ffrel,ps1,oneloop2}. The argument above extends this result 
to hadrons of arbitrary spin and multibody states $M_n X_S$.

Another prediction is a relation between the time-like and longitudinal 
nonfactorizable contributions to the helicity amplitudes for an arbitrary
current $\Gamma$ 
\begin{eqnarray}\label{hel0t}
\frac{H^{\rm nf}_t(B\to M_n X_S)}{H^{\rm nf}_0(B\to M_n X_S)} &=& 
 \frac{c_2 (v\cdot \varepsilon_t^*) + c_3 (n\cdot \varepsilon_t^*)}
{c_2 (v\cdot \varepsilon_0^*) + c_3 (n\cdot \varepsilon_0^*)}\\
&+& O(\frac{\lqcd}{m_b})\nonumber
\end{eqnarray}

Finally, SCET predicts also the ratio of helicity amplitudes mediated by
different currents,
into any state $M_n X_S$ containing one energetic
collinear particle, e.g.
\begin{eqnarray}\label{HVA}
&& \frac{H_-^{V-A}(B\to M_n X_S)}{H_-^{T}(B\to M_n X_S)} = 
\frac{c_1^{(V-A)}(E_M)}{c_1^{(T)}(E_M)}\\
&& \hspace{3cm} + O(\frac{\lqcd}{m_b}) \nonumber \\
&& \frac{H_0^{V-A}(B\to M_n X_S)}{H_0^{T}(B\to M_n X_S)} = \\ 
&&\quad \frac{c_2^{(V-A)} (v\cdot \varepsilon_0^*) + c_3^{(V-A)} (n\cdot \varepsilon_0^*)}
{c_2^{(T)} (v\cdot \varepsilon_0^*) + c_3^{(T)} (n\cdot \varepsilon_0^*)}
+ O(\frac{\lqcd}{m_b})\nn
\end{eqnarray}

These relations are in general broken by the factorizable contributions from
Eq.~(\ref{fact2}). For example, the helicity zeros (\ref{hel0}) could disappear
if the $b_{1R}$ term gives a nonvanishing contribution (note that 
the $b_{1R(L)}$ term in Eq.~(\ref{fact2}) contributes only to the $H_{+(-)}$
helicity amplitude). For a 1-body state, 
this is forbidden by angular momentum conservation since the collinear
part of the operator can only produce a longitudinally polarized meson. 
{\em However, this constraint does not apply for multibody final states 
$M_n X_S$} (except in channels of well defined $J^P$ quantum numbers). In particular, 
this means that the helicity zero Eq.~(\ref{hel0}) receives corrections at leading order 
in $1/m_b$. These corrections are computed in Sec.~4.

The factorizable corrections to these relations 
are parameterized in terms of the soft functions
\begin{eqnarray}\label{Smu}
&&S^{(R)}(k_+,S_X) = \langle X_S|
(\bar q Y)_{k_+} \nslash \varepsslash_+^* P_R (Y^\dagger b_v) |\bar B\rangle\\
&&S^{(L)}_{\lambda}(k_+,S_X) = \langle X_S|
(\bar q Y)_{k_+} \nslash \varepsslash_-^* \gamma^\perp_\lambda P_R 
(Y^\dagger b_v)|\bar B\rangle\nn\\
&& \hspace{3cm}  \equiv - \frac12 S^{(L)}(k_+, X_S) \varepsilon_+^\lambda  
\nn\\
&&S^{(0)}(k_+,S_X) = \langle X_S|
(\bar q Y)_{k_+} \nslash P_L 
(Y^\dagger b_v)|\bar B\rangle\nn
\end{eqnarray}
Parity invariance of the strong interactions gives one relation among
these functions in channels with $X_S^{J\Pi}$ of well-defined spin $J$ 
and intrinsic parity $(-)^\Pi$
\begin{eqnarray}
S^{(L)}(k_+, S_X^{J\Pi}) &=& \langle X_S^{J\Pi}|
(\bar q Y)_{k_+} \nslash P_R (Y^\dagger b_v)|\bar B\rangle \nn\\
&=&  (-)^{J+\Pi-1} S^{(0)}(k_+, \hat R_\pi \hat P S_X^{J\Pi})
\end{eqnarray}
where $\hat P$ is the parity operator and $\hat R_\pi$ the rotation
operator by $180^\circ$ around the $y$ axis.

Compared with the decays into one-body hadronic states, for which only
the soft function $S^{(0)}$ is required, this represents an increase in the
number of independent parameters. However, the total number is still less
than in QCD (see Table 1), such that predictive power is retained.
In the next section we construct a chiral formalism which can be used 
to compute these matrix elements for any state $X_S$ containing only soft pions.

\vspace{0.5cm}

3. \textit{Chiral formalism.} We construct here the representation of the
soft operator $O_S$ giving the soft functions in (\ref{Smu}) 
in the low energy chiral theory. Since we are interested in
B decays, the appropriate tool is the heavy hadron chiral perturbation theory
developed in Refs.~\cite{hhchpt}. The main result is that the matrix elements
of $O_S$ depend only on the B meson light cone wave function.

The effective Lagrangian that describes the
low momentum interactions of the $B$ mesons
with the pseudo-Goldstone bosons $\pi, K$ and $\eta$ is invariant
under chiral $SU(3)_L \times SU(3)_R$ symmetry and
under heavy quark spin symmetry.  This requires the introduction
of the heavy quark doublet $(B, B^*)$ as the relevant matter field.
The chiral Lagrangian for matter fields such as the $B^{(*)}$ 
must be written in terms
of velocity dependent fields, to preserve the validity of the chiral
expansion. 

The chiral effective Lagrangian describing 
the  ground state mesons
containing a heavy quark $Q$ is \cite{hhchpt}
\begin{eqnarray}\label{Lag}
{\cal L} &=& {f^2 \over 8}Tr 
\left( \partial^{\mu} \Sigma
\partial_{\mu} \Sigma^\dagger \right)
+\lambda_0 Tr\ \left[ m_q \Sigma + m_q \Sigma^\dagger \right]\\
& &
-i Tr \bar H^{(Q)a} v_{\mu} \partial^{\mu} H_a^{(Q)} \nonumber \\
& &+{i \over 2} Tr \bar H^{(Q)a} H_b^{(Q)} v^{\mu} \left[ \xi^\dagger
\partial_{\mu} \xi + \xi \partial_{\mu} \xi^\dagger \right]_{ba} \nonumber\\
& &+{{ig} \over 2} Tr \bar H^{(Q)a} H_b^{(Q)} \gamma_{\nu} \gamma_5
\left[\xi^\dagger \partial^{\nu} \xi - \xi \partial^{\nu}
\xi^\dagger \right]_{ba} + \cdots \nonumber 
\end{eqnarray}
where the ellipsis denote light quark mass terms, $O(1/m_b)$
operators associated with the breaking of heavy quark spin symmetry,
and terms of higher order in the derivative expansion.
The pseudoscalar and vector
heavy meson fields $P_a^{(Q)}$ and $P^{*(Q)}_{a\mu}$ form the matrix 
\begin{eqnarray}
H_a^{(Q)} = \frac{1+\vslash }{2} \left[ P^{*(Q)}_{a \mu} \gamma^{\mu}
- P_a^{(Q)} \gamma_5 \right]. 
\end{eqnarray}
For $Q=b$, $(P_1^{(b)},P_2^{(b)},P_3^{(b)})=
(B^-, \bar B^0, \bar B_s)$, and similarly for $P^{*(b)}_{a\mu}$.
The field $H^{(Q)}_a$ transforms as a
$\bar 3$ under flavor $SU(3)_V$,
\begin{eqnarray}
H^{(Q)}_a \rightarrow   H^{(Q)}_b \ U^\dagger_{ba} .
\end{eqnarray}
and describes $\bar B$ and $\bar B^*$ mesons with definite velocity
$v$. For simplicity of notation we will omit the subscript $v$ on $H$, $P$ and $P^*_\mu$.
The pseudo-Goldstone bosons appear in the Lagrangian through
$\xi = e^{iM/f}$ ($\Sigma=\xi^2$) where
\begin{eqnarray}
M = \left(
\begin{array}{ccc}
 {1\over\sqrt2}\pi^0 +
{1\over\sqrt6}\eta &
\pi^+ & K^+ \\
\pi^-& -{1\over\sqrt2}\pi^0 + {1\over\sqrt6}\eta&K^0 \\
K^- &\bar K^0 &- {2\over\sqrt6}\eta \\
\end{array}
\right)
\end{eqnarray}
and the pion decay constant $f \simeq 135$~MeV. These fields transform as
\begin{eqnarray}
\Sigma \to L \Sigma R^\dagger\,,\qquad \xi \to L \xi U^\dagger = U \xi R^\dagger
\end{eqnarray}
The Lagrangian Eq.~(\ref{Lag})
is the most general Lagrangian invariant under both the heavy quark and
chiral symmetries to leading order in $m_q$ and $1/m_Q$.

The symmetries of the theory constrain also the form of operators such as currents.
For example, the left handed current
$L^{\nu}_a = \bar q_a \gamma^\nu P_L Q$ in QCD can be written
in the low energy chiral theory as \cite{hhchpt}
\begin{eqnarray}\label{JL}
L^{\nu}_a = \frac{i \alpha}{2}
\mbox{Tr} [\gamma^\nu P_L H_b^{(Q)} \xi^\dagger_{ba}] + ...,
\end{eqnarray}
where the ellipsis denote higher dimension operators in the chiral and
heavy quark expansions. The parameter $\alpha$ is obtained by taking the
vacuum to B matrix element of the current, which gives
$\alpha=f_B\sqrt{m_B}$ (we use a nonrelativistic normalization for
the $|B^{(*)}\rangle$ states as in \cite{hhchpt}).

In the SCET we require also the matrix elements
of nonlocal operators $O_S$, which appear in Eq.~(\ref{fact}). 
To leading order in $1/m_b$ these operators are quark bilinears
\begin{eqnarray}
& &O^a_{L,R}(k_+) = \\
& &\quad  \int \frac{d x_-}{4\pi} e^{-\frac{i}{2} k_+ x_-}
\bar q^a(x_-) Y_n(x_-,0) P_{R,L} \Gamma b_v(0)\,.\nn
\end{eqnarray}
Under the chiral group they transform as
$(\mathbf{\overline{3}}_L, \mathbf{1}_R)$ and
$(\mathbf{1}_L, \mathbf{\overline{3}}_R)$.
In analogy with the local current (\ref{JL}) we write for $O^a_{L,R}(k_+)$
in the chiral theory
\begin{eqnarray}\label{JOL}
&& O^a_{L}(k_+)  = \frac{i}{4} 
\mbox{Tr} [\hat \alpha_L(k_+) P_R\Gamma H_b^{(Q)} \xi^{\dagger }_{ba}],\\
\label{JOR}
&& O^a_{R}(k_+)  = \frac{i}{4} 
\mbox{Tr} [\hat \alpha_R(k_+) P_L\Gamma H_b^{(Q)} \xi_{ba}]
\end{eqnarray}
where the most general form for $\hat \alpha_{L,R}(k_+)$ depends on eight
unknown functions $a_i(k_+)$
\begin{eqnarray}
\hat \alpha_{L,R}(k_+) = a_{1L,R} + a_{2L,R} \nslash + a_{3L,R} \vslash +
\frac12 a_{4L,R} [\nslash, \vslash]
\end{eqnarray}
The heavy quark symmetry constraint $H^{(Q)} \vslash = - H^{(Q)}$ reduces the 
number of these functions to four. Taking the vacuum to B meson matrix element 
fixes the remaining functions as
\begin{eqnarray}\label{alpha}
\hat\alpha_L(k_+) =\hat\alpha_R(k_+) = f_B \sqrt{m_B} [\bnslash \phi_+(k_+) + 
\nslash \phi_-(k_+)]
\end{eqnarray}
where $\phi_\pm(k_+)$ are the usual light-cone wave functions of a B meson, 
defined by \cite{wv}
\begin{eqnarray}
&& \int \frac{dz_- }{4\pi} e^{-\frac{i}{2} k_+ z_-}
\langle 0 | \bar q_i(z_-)Y_n(z_-,0) b_v^j(0)|\bar B(v)\rangle = \\
&& -\frac{i}{4}f_B \sqrt{m_B} \left\{
\frac{1+\vslash}{2} [\bnslash n\cdot v \phi_+(k_+) +
\nslash \bn\cdot v \phi_-(k_+) ]\gamma_5 \right\}_{ij} \nonumber
\end{eqnarray}
We find thus the remarkable result that the B meson light-cone wave functions
are sufficient to fix the pion matrix elements of the nonlocal operators
$O_{L,R}^a(k_+)$.

The same result can be obtained also by considering only local operators. 
Let us consider the operator $O_L^a(k_+)$ (the same results are obtained
for $O_R^a(k_+)$). Expanding in a power series of the distance 
along the light cone one is led to consider the matrix elements of the operator
symmetric and traceless in its indices 
\begin{eqnarray}\label{On}
O_{L}^{a,\mu_1\mu_2\cdots \mu_N} &=& \bar q^a (-i \Dleft)^{\{ \mu_1}
\cdots (-i \Dleft)^{\mu_N \}} P_{R} \Gamma b_v\, \\
& & - (\mbox{traces})\nn
\end{eqnarray}
Heavy quark and chiral symmetry constrain the chiral effective representation
of this operator to be of the form
\begin{eqnarray}
O_{L}^{a,\mu_1\mu_2\cdots \mu_N} =
\frac{i}{4} 
\mbox{Tr} [\sum_j \alpha_{N,j} X_j^{\mu_1\mu_2\cdots \mu_N} P_R\Gamma H_b^{(Q)} \xi^{\dagger }_{ba}],
\end{eqnarray}
where the sum over $j$ includes the most general symmetric and traceless
structures $X$ formed from $\gamma_\mu, v_\mu, g_{\mu\nu}$. There are many 
such structures, but only 2 of them survive when contracted with 
$n_{\mu_1}\cdots n_{\mu_N}$
\begin{eqnarray}
&& X_0^{\mu_1\mu_2\cdots \mu_N} = v^{\mu_1}v^{\mu_2} \cdots v^{\mu_N} - 
(g^{\mu_i \mu_j}\mbox{-terms}) \\
&& X_1^{\mu_1\mu_2\cdots \mu_N} = \gamma^{\{\mu_1}v^{\mu_2} \cdots v^{\mu_N\}} 
- (g^{\mu_i \mu_j}\mbox{-terms})\nn 
\end{eqnarray}
This gives the chiral representation of the projection of the operators 
(\ref{On}) on the light-cone
\begin{eqnarray}
&& \bar q^a (-i n\cdot \Dleft)^N P_{R} \Gamma b_v \\
&& \qquad \to
\frac{i}{4} 
\mbox{Tr} [(\alpha_{N,0} + \nslash \alpha_{N,1})  
P_R\Gamma H_b^{(Q)} \xi^{\dagger }_{ba}],\nn
\end{eqnarray}
which makes it clear that the constants $\alpha_{N,0}, \alpha_{N,1}$ are 
uniquely fixed in terms of the $B\to $ vacuum matrix elements of the
operators (\ref{On}). Assuming that the B light-cone wave functions
are well behaved at large $k_+$, these matrix elements are related
to the moments of $\phi_\pm(k_+)$.
Specifically, one finds
\begin{eqnarray}
\alpha_{N,0} &=& -2f_B \sqrt{m_B}
\int \mbox{d} k_+ (k_+)^N \phi_+(k_+)\,.
\end{eqnarray}
In particular, for $N=1$ this gives $\alpha_{1,0}=-\frac83 f_B \sqrt{m_B} \bar\Lambda$, 
which agrees with Ref.~\cite{GrPi}. 

Beyond leading order in $1/m_b$ many more operators can be written.
For example, the matrix elements of $O_L^a$ with 
one insertion of the chromomagnetic term in the HQET Lagrangian
${\cal L}_m = g\bar b_v \sigma_{\mu\nu} G^{\mu\nu} b_v$
gives structures of the form
\begin{eqnarray}
T\{O_L^a, i{\cal L}_m\}\to \mbox{Tr}[X^{\mu\nu}
P_R \Gamma \frac{1+\vslash}{2}i\sigma_{\mu\nu} H_b^{(Q)} \xi^\dagger_{ba} ]
\end{eqnarray}
with $X^{\mu\nu} = \beta_1 [n^\mu, \gamma^\nu] + \beta_2 i\sigma^{\mu\nu} +
\nslash \beta_3 [n^\mu, \gamma^\nu] + \beta_4 \nslash i\sigma^{\mu\nu} $.
The proliferation of unknown constants (see also \cite{GrBo}) spoils the simple 
leading order result that knowledge of the $B\to$ vacuum matrix element is 
sufficient to fix all low energy constants.

The operators in Eqs.~(\ref{JOL}), (\ref{JOR}) (together with (\ref{alpha}))
give the desired representation of the soft operators $O_{L,R}$ in the chiral
effective theory, and can be used to compute their matrix elements 
on states with a $B$ meson and any number of pseudo Goldstone bosons.
\vspace{0.5cm}

\begin{figure}
\includegraphics[height=1.5cm]{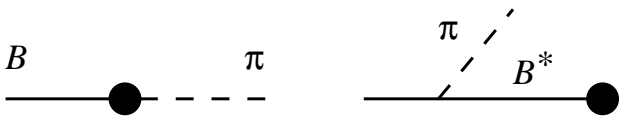}%
\caption{\label{fig1} Feynman diagrams for $\bar B \to M_n \pi $.
The filled circle denotes one insertion of the soft operator $O_S$. The 
collinear hadron $M_n$ is not shown.}
\end{figure}

4. \textit{Application: $\bar B\to M_n \pi\ell\bar\nu$ and 
$\bar B\to M_n \pi\ell^+\ell^-$.}
As an application we compute the
factorizable corrections to the symmetry relations (\ref{hel0}), (\ref{HVA}) 
for the transverse helicity amplitudes in $\bar B\to M_n \pi\ell\bar\nu$
in the region of the phase space with one energetic meson 
$M_n = \pi, \rho, K^*,$ etc. plus one soft pion.

The factorizable contribution to the transverse helicity amplitudes 
for $B\to M_n \pi$ are given by the matrix elements of Eq.~(\ref{fact2}).
Specifically, one has for $M_n$ a pseudoscalar meson
\begin{eqnarray}\label{amp1P}
&& H^{\rm fact}_{+}(\bar B\to P_n(k) X_S) = \\
&& \hspace{2cm}  C \frac18 f_B f_P m_B S_{R}(X_S)
\langle b_{1R} J_\parallel \phi_P \rangle  \nn \\
\label{H-2P}
&& H^{\rm fact}_{-}(\bar B\to P_n(k) X_S) = 0
\end{eqnarray}
and for $M_n$ a vector meson
\begin{eqnarray}\label{amp1}
&& H^{\rm fact}_{+}(\bar B\to V_n(k,\eta) X_S) = \\
&& \qquad  C \frac{f_B f_V m_B m_V}{8\bar n\cdot p_V} S_{R}(X_S) (\bn\cdot \eta^*) 
\langle b_{1R} J_\parallel \phi_V^\parallel \rangle \nn \\
\label{H-2}
&& H^{\rm fact}_{-}(\bar B\to V_n(k,\eta) X_S) = \\
&& \qquad -\frac14 C f_B f_V^\perp m_B
 S_{L}(X_S) (\varepsilon_-^*\cdot \eta^*)
\langle b_{1L} J_\perp \phi^\perp_V \rangle \nn
\end{eqnarray}
We used here the short notation $\langle b_i J_a \phi^a \rangle =
\int dx dz dk_+  b_i(x,z) J_a(x,z,k_+) \phi_+(k_+) 
\phi^a(x)$.
The isospin factor $C$ depends on the collinear meson, e.g. $C(\rho^0)=1/\sqrt2,
C(\rho^\pm) = 1$. 
The corresponding results for the 1-body factorizable
decay amplitudes are obtained from these expressions by taking
$S_R \to 0\,, S_L \to 1$.

Inspection of the results (\ref{amp1P})-(\ref{H-2}) gives the 
following conclusions, valid to all orders in $\alpha_s$. 

i) The null result in Eq.~(\ref{H-2P}) means
that the symmetry relation (\ref{HVA}) for $\bar B\to P_n X_S$ transitions
is not broken by factorizable corrections and {\em
is thus exact to leading order in $1/m_b$}.
This leads, {\em e.g.,} to a relation between $\bar B\to 
(\bar K_n\pi_S)_{h=-1} e^+e^-$
and $\bar B\to (\pi_n \pi_S )_{h=-1} e^-\bar\nu$.

ii) The vanishing of the $H_{+}$ nonfactorizable
helicity amplitudes in $\bar B$ decay
Eq.~(\ref{hel0}) is violated by the factorizable terms Eqs.~(\ref{amp1P}),
(\ref{amp1}). These terms are however calculable in chiral perturbation
theory for $X_S$ containing only soft pions. For both $M_n = P,V$, the pion 
carries $m_3=+1$ angular momentum; the $V_n$ collinear meson is emitted longitudinally 
polarized.


The soft functions $S_{R,L}(p_\pi)$ in Eq.~(\ref{amp1P}), (\ref{H-2})
can be computed explicitly in terms of the
chiral perturbation theory diagrams in Fig.~1. We find
\begin{eqnarray}
&& S_R(p_\pi) = \frac{g}{f_\pi} 
\frac{\varepsilon_+^*\cdot p_\pi}{v\cdot p_\pi + \Delta - i\Gamma_{B^*}/2} \\
\label{SLchpt}
&& S_L(p_\pi) = \frac{1}{f_\pi}\left(
1  -  g \frac{e_3 \cdot p_\pi}
{v\cdot p_\pi + \Delta - i\Gamma_{B^*}/2}\right)
\end{eqnarray}
with $\Delta = m_{B^*} - m_B \simeq 50$ MeV and $\Gamma_{B^*}$ the width of the $B^*$ meson.
While the soft matrix elements in Eq.~(\ref{Smu}) have a factorized
form $S^{(i)}(k_+,S_X) = \phi_+(k_+) S_i(S_X)$, the total factorizable
amplitude is not simply the product $B\to B^*\pi$ times $B^*\to  M_n$,
due to the direct graph in Fig.~1a (nonvanishing only for $S_L$). 
At threshold, 
the relation Eq.~(\ref{SLchpt}) gives a soft pion theorem which fixes the
soft function in $B\to M_n \pi$ in terms of the factorizable contribution 
to the $B\to M_n $ transition.
Note that the $B^*$ width in the propagator is a 
source of strong phases at leading order in $1/m_b$.
These  results can be extended to final states containing multiple
soft pions, without introducing any new unknown hadronic parameters. 

\vspace{0.5cm}

5. \textit{Conclusions.}
We presented in this paper the application of the soft-collinear effective
theory to B decays into multibody final states, containing one energetic
meson plus soft pseudo Goldstone bosons. The additional ingredient is the
application of heavy hadron chiral perturbation theory \cite{hhchpt} to compute 
the matrix elements with Goldstone bosons of the nonlocal soft operators obtained 
after factorization. (This assumes that the only SCET operators contributing to
these decays are the same as those describing $B\to M_n$ transitions \cite{bpsff}.)
Heavy quark and chiral 
symmetry are powerful constraints which fix all these couplings in terms
of the usual B light-cone wave functions. This simplicity should be contrasted
with the case of the twist-2 DIS and DVCS operators, whose matrix elements
on nucleons plus soft pions 
require additional couplings not constrained by the nucleon structure
functions \cite{ChSa}.

Some of the symmetry predictions of SCET rely on angular momentum conservation
arguments which are invalidated when the final hadronic state contains more than
one hadron (see Eq.~(\ref{hel0})), already at leading order in the $1/m_b$ 
expansion. The chiral
formalism presented here allows the systematic computation of these effects.
We point out the existence of an exact relation Eq.~(\ref{HVA}) among 
left-handed helicity amplitudes in $\bar B\to P_n X_S$ transitions
induced by different $b\to q_n$ currents.

These results extend the applicability of SCET to B decays into multibody states
$M_n X_S$ containing one energetic particle. 
It is interesting to note that
the corrections to these predictions 
scale like $\mbox{max}(\Lambda/E_M\,, p_S/\Lambda_{\chi pT})$,
rather than $m_X/E_M$. This suggests that the range of validity of 
factorization in these decays might be wider than previously
thought, a fact noted empirically in 
Refs.~\cite{DX} in the context of the $B\to DX$ decays.
Many more problems can be studied using the formalism described here, 
e.g., the leading SU(3) violating contributions to the factorizable 
contributions, analogous to the effects considered in Ref.~\cite{parton1,fB}.



\begin{acknowledgments}
We are grateful to Martin Savage for useful discussions. 
B.G. was supported in part by the DOE under Grant DE-FG03-97ER40546.
D.P. was supported by the
U.S.\ Department of Energy under cooperative research agreement 
DOE-FC02-94ER40818.
\end{acknowledgments}


\end{document}